\documentclass[12pt]{article}

\usepackage{amsmath,amssymb}
\usepackage{graphicx}

\makeatletter

\renewcommand\section{\@startsection{section}{1}{\z@}
                                   {-3.5ex \@plus -1ex \@minus -.2ex}
                                   {2.3ex \@plus .2ex}
                                   {\normalfont\large\bfseries}}
\renewcommand\subsection{\@startsection{subsection}{2}{\z@}
                                   {-3.25ex\@plus -1ex \@minus -.2ex}
                                   {1.5ex \@plus .2ex}
                                   {\normalfont\normalsize\bfseries}}
\renewcommand\subsubsection{\@startsection{subsubsection}{3}{\z@}
                                   {-3.25ex\@plus -1ex \@minus -.2ex}
                                   {1.5ex \@plus .2ex}
                                   {\normalfont\normalsize\bfseries}}
\renewcommand\paragraph{\@startsection{paragraph}{4}{\z@}
                                   {3.25ex \@plus1ex \@minus.2ex}
                                   {-1em}
                                   {\normalfont\normalsize\bfseries}}

\makeatother

\newcommand{\be}{\begin{equation}}
\newcommand{\ee}{\end{equation}}
\newcommand{\bea}{\begin{eqnarray}}
\newcommand{\eea}{\end{eqnarray}}
\newcommand{\ba}{\begin{array}}
\newcommand{\ea}{\end{array}}

\newcommand{\id}{\hbox{1\kern-.27em l}}
\newcommand{\ZZ}{\mathbb{Z}}
\newcommand{\RR}{\mathbb{R}}
\newcommand{\half}{ {\textstyle \frac{1}{2}  } }

\newcommand{\al}{\alpha}

\newcommand{\ep}{\epsilon}
\newcommand{\si}{\sigma}

\newcommand{\cN}{\mathcal{N}}

\newcommand{\cW}{\mathcal{W}}

\newcommand{\tr}{{\rm tr}}

\newcommand{\rar}{\rightarrow}
\newcommand{\non}{\nonumber}

\newcommand{\SU}{\mathrm{SU}}
\newcommand{\SO}{\mathrm{SO}}
\newcommand{\SL}{\mathrm{SL}}
\newcommand{\Sp}{\mathrm{Sp}}
\newcommand{\su}{\mathrm{su}}
\newcommand{\so}{\mathrm{so}}
\newcommand{\spl}{\mathrm{sp}}

\newcommand{\ul}{\mathrm{u}}

\begin{document}

\begin{center}

\vspace*{5mm}
{\Large\sf   A note on S-duality for the $\cN=1^*$ $\Sp(2n)$ \\ 
and $\SO(2n{+}1)$ super-Yang-Mills theories }

\vspace*{5mm}
{\large Niclas Wyllard}

\vspace*{5mm}
Department of Fundamental Physics\\
Chalmers University of Technology\\
S-412 96 G\"oteborg, Sweden\\[3mm]
{\tt wyllard@fy.chalmers.se}          

\vspace*{5mm}{\bf Abstract:} 

\end{center}

\noindent We study the $\cN=1^*$ supersymmetric gauge theories with gauge groups $\Sp(2n)$ and $\SO(2n{+}1)$. These theories are obtained from the corresponding $\cN=4$ supersymmetric Yang-Mills theories via a mass deformation. We show that the number of quantum vacua in the $\Sp(2n)$ theory is equal to the number of quantum vacua in the $\SO(2n{+}1)$ theory. This constitutes non-trivial support for S-duality between these theories. The verification of the equality of the number of quantum vacua involves a rather esoteric identity due to Ramanujan.

\setcounter{equation}{0}
\section{Introduction}
The $S$-duality conjecture \cite{Montonen:1977} for the  $\cN = 4$ supersymmetric four-dimensional Yang-Mills theories is the statement   that the theory with gauge group $G$ and a value of the complex parameter
$
\tau = \frac{\theta}{2 \pi} + \frac{i}{g_{\rm YM}^2},
$
where $\theta$ is the theta angle and $g_{\rm YM}$ is the coupling constant, is equivalent to the theories arising from the transformations $S$ and $T$: 
\bea \label{S}
S : (G, \tau) & \rightarrow & ( G^{\vee}, - 1 / r \tau) \cr
T : (G, \tau) & \rightarrow & (G, \tau + 1) \,,
\eea
where $G^{\vee}$ denotes the dual group of $G$ \cite{Goddard:1976} and $r$ is the square of the ratio of the long and short roots of the Lie algebra of $G$ (see e.g.~\cite{Argyres:2006} for a recent discussion). For the simple groups with simply-laced Lie algebras, $G^{\vee}$ and $G$ are equal at the Lie algebra level. However, this is not true for all groups. For instance, for $G=\Sp(2n)$ the dual is $G^{\vee}=\SO(2n{+}1)$. 

In this note we study the so called $\cN=1^*$ gauge theories, obtained by adding a mass deformation to the corresponding  $\cN=4$ Yang-Mills theories. It is believed that S-duality (\ref{S}) is inherited from the $\cN=4$ models and therefore also realised in the $\cN=1^*$ theories. 

The quantum vacuum structure of the $\SU(n)$ $\cN=1^*$ theory was elucidated in~\cite{Dorey:1999} (see \cite{Vafa:1994} for earlier work on the vacuum structure). In \cite{Dorey:1999} it was shown that the quantum vacua are controlled by a superpotential which coincides with the potential of the (complexified) elliptic Calogero-Moser model. The stationary points of this potential are known, which, in particular, makes it possible to explicitly describe the action of the $\SL(2,\ZZ)$ S-duality group on the (massive) vacua. 

For the $\cN=1^*$ theories based on the other simple groups there exist conjectured superpotentials~\cite{Kumar:2001} which should give the quantum vacuum structure. Also for the other groups the superpotentials are given by potentials of (twisted) elliptic Calogero-Moser models. Unfortunately, the stationary points of these models are not known, and therefore the same analysis as for $\SU(n)$ can not be done. 
One possible way out of this impasse would be to numerically look for stationary points that lie on some lattice (as presumably would be required for $S$-duality to work).  However, unless exact expressions can be found, such an approach, even if it works, would at best be a method that could be applied for groups with low ranks. 

Rather than to look for exact expressions for the stationary points of the conjectured superpotentials of \cite{Kumar:2001}, in this note we have a more modest goal. Since the $\cN=1^*$ theories with gauge groups $\Sp(2n)$ and $\SO(2n{+}1)$ are related by the $S$ transformation of the S-duality group they should, in particular, have the same number of quantum vacua (as the number of vacua is independent of the coupling constant). The purpose of this note is to check whether the number of vacua agrees for the $\Sp(2n)$ and $\SO(2n{+}1)$  $\cN=1^*$  theories.

Let us also mention that in a recent paper \cite{Henningson:2007} S-duality was investigated for another variant of the $\cN=4$ theories, namely the $\cN=4$ Yang-Mills theories on the space $\RR{\times}T^3$. It should be possible to combine the results here and in \cite{Henningson:2007}, i.e.~to study the $\cN=1^*$ theories on $\RR{\times}T^3$. However,  the two setups are in a sense orthogonal and combining them does not seem to lead to a richer structure. 

In the next section we review the (quantum) vacuum structure of the $\Sp(2n)$ and $\SO(2n{+}1)$ theories, and then in section 3 we determine the number of quantum vacua in the two theories and check that they are equal, as required by S-duality.

\setcounter{equation}{0}
\section{Vacuum structure of the $\cN=1^*$ gauge theories}

In this section, we review the 
vacuum structure of the mass-deformed $\cN=4$ Yang-MIlls theories 
known as the $\cN = 1^*$ theories~\cite{Vafa:1994}, focusing on the gauge groups 
$\SO(n)$ and $\Sp(2n)$. 
The superpotential of the $\cN=4$ model, 
written in terms of $\cN=1$ superfields, is
\be \label{W} \cW =
\frac{2\sqrt{2}}{6g_{\mathrm{YM}}^2}\ep^{ijk} \tr
(\phi^i[\phi^j,\phi^k])\,,
\ee
where $\phi^i$  ($i = 1,2,3$) are  chiral superfields
transforming in the adjoint representation of the gauge group.
For $\SO(n)$, the $n \times n$ matrices $\phi^i$ must satisfy 
 \be 
 \label{sodef} 
 (\phi^i)^T  =  -\phi^i \,, 
 \ee
appropriate  to the generators of the adjoint representation of $\so(n)$.
For $\Sp(2n)$,  
the $2n\times 2n$ matrices $\phi^i$ must satisfy 
 \be 
 \label{spdef} 
 J (\phi^i)^T J =  \phi^i \,,
 \ee
which defines the generators of the adjoint representation of $\spl(2n)$.
The real matrix $J$ is the symplectic unit of $\Sp(2n)$, 
satisfying $J^T= -J$ and $J^2=-\id_{2n}$.

To obtain the $\cN=1^*$ theory one adds to the superpotential (\ref{W})  
 the mass deformation 
\be 
\label{DW}  
\cW_{\rm mass} =
\frac{\sqrt{2}}{g_{\mathrm{YM}}^2}\sum^3_{i=1} m_i \tr(\phi^i)^2\,,
\ee
which, when all the masses are non-zero,
breaks the supersymmetry down to $\cN=1$.
In what follows, we rescale the $\phi^i$ to make the masses equal 
to one. (This rescaling does not affect the vacuum structure.) 
The classical supersymmetric vacuum states 
are obtained by solving the F- and D-term equations, 
which are 
 \be
 \label{F} 
 {}[\phi^i,\phi^j] = - \ep^{ijk}\phi^k \,,
 \ee
and
 \be 
 \label{D} 
 \sum_{i=1}^3 [\phi^i,(\phi^i)^{\dagger}]=0 \,,
 \ee
respectively. 
Equation (\ref{F}) together with (\ref{D}) 
imply that the $\phi^i$ are anti-hermitian \cite{Kac:1999b}.
Furthermore, (\ref{F}) imply that the $\phi^i$ form a  
(in general reducible) representation 
of the $\su(2)$ Lie algebra. 
It is always possible to choose a block-diagonal basis,
 \be 
 \label{Ti}
 \phi^i = \left( \ba{ccc} T^i_{n_1}  & & \\ 
                        & \ddots &   \\
                        & & T^i_{n_l}
 \ea \right ),  
 \ee 
in which $T^i_{n_k}$ are the generators of the $n_k$-dimensional
irreducible representation of $\su(2)$. For $\SO(n)$ ($\Sp(2n)$) 
$\sum_{k=1}^{l} n_k$ equals $n$ ($2n$).
For the gauge group $\SU(n)$ the above argument gives the complete solution, but for 
 $\SO(n)$ and $\Sp(2n)$ the  conditions on the $\phi^i$'s, (\ref{sodef}) and (\ref{spdef}), lead to restrictions on the allowed dimensions of the $\su(2)$ irreps. (Note that when choosing the block-diagonal form (\ref{Ti}) for the $\phi^i$, the form of the conditions (\ref{sodef}) and (\ref{spdef}) in general change).  The restrictions  were worked out in~\cite{Naculich:2001} and are summarised in the following table:

\begin{center} 
\begin{tabular}{|c|c|c|} 
\hline
Gauge group & Allowed $\su(2)$ representations & gauge enhancement \\ 
\hline
 & &  \\[-13pt]
\hline 
 & &  \\[-13pt]
$\Sp(2n)$   & $2m$ odd-dimensional irreps 
          & $\spl(2m)$ \\[1pt]
\cline{2-3}
 & &  \\[-13pt]
{} & $m$ even-dimensional irreps & $\so(m)$  \\
 & &  \\[-13pt]
\hline
& &  \\[-13pt]
\hline
& &  \\[-13pt]
$\SO(n)$  & $2m$ even-dimensional irreps 
           & $\spl(2m)$  \\[1pt]
\cline{2-3}
& &  \\[-13pt]
{} & $m$ odd-dimensional irreps & $\so(m)$ \\ 
\hline
\end{tabular}  
\end{center} 

The building blocks given in this table can be used to construct the  complete solution to the classical vacuum 
problem. In general, the vacuum breaks the gauge symmetry down to a subgroup. 
The form of this group was also derived in \cite{Naculich:2001} and is summarised in the above table.

Before we continue, let us comment on an at first sight puzzling aspect 
of the above result.  
Dynkin has classified the number of ways an $\su(2)$ subalgebra can be 
embedded into a simple Lie algebra \cite{Dynkin:1957}. This classification seemingly differs 
from the above result. For instance, for $\Sp(4)$ we find four vacua, namely (in terms of the dimensions of the $\su(2)$ representations)  $\mathbf{4}$, $\mathbf{2+2}$, $\mathbf{2+1+1}$ and $\mathbf{1+1+1+1}$, whereas the result in \cite{Dynkin:1957} gives five solutions to the general $\su(2)$ embedding problem. Our understanding of this discrepancy is that the result in \cite{Dynkin:1957} is a representation independent statement and that in certain representations (as is the case here and in \cite{Naculich:2001}) some of the solutions may be isomorphic but in general this does not occur. For instance, for $\Sp(4)$, one can explicitly check that out of the five possibilities appearing in the table in \cite{Dynkin:1957}, in our setup two are actually isomorphic, i.e.~can be transformed into each other \cite{Naculich:2002}.

Above we reviewed the classification of the classical vacua.  Near each classical vacuum the theory is described by an $\cN=1$ supersymmetric Yang-Mills theory with gauge group $h$, where $h$ is the unbroken gauge symmetry in that vacua. It is well known that the number of quantum vacua in such a theory is given by $g^\vee_h$, the dual Coxeter number of $h$. (Recall that $g^\vee_h$ is $m{+}1$ for $\spl(2m)$ and $m{-}2$ for $\so(m)$ ($m>4$).) 
Combining this result with the above classical analysis we obtain an algorithm which can be used to determine the number of quantum vacua.

\setcounter{equation}{0}
\section{S-duality for the $\Sp(2n)$ and $\SO(2n{+}1)$  $\cN=1^*$ theories}

In the previous section we discussed the (quantum) vacuum structure of the $\Sp(2n)$ and $\SO(2n{+}1)$  $\cN=1^*$ theories. As an example we list the result for $\SO(7)$:
 \be \label{so7}
\begin{tabular}{ccc}
\!\! \underline{$\su(2)$ irreps} & \!\!\!\!\!\!\!\! \!\!\!\! \underline{unbroken symmetry} & \underline{number of quantum vacua}  \\
\!\! $\mathbf{7}$ &\!\!\!\!\!\!\!  \!\!\!\!\!$\emptyset$ &  $1$  \\
\!\! $\mathbf{5+1+1}$ &\!\!\!\!\!\!\! \!\!\!\!\! $\so(2)$ & $\mathrm{massless }$ $(1)$  \\
\!\! $\mathbf{3+3+1}$ &  \!\!\!\!\!\!\!  \!\!\!\! $\so(2)$ & $\mathrm{massless }$ $(1)$  \\
\!\!$\mathbf{3+2+2}$ &\!\!\!\!\!\!\!   \!\!\!\!$\spl(2)$ & $2$  \\
\!\!$\mathbf{3+1+1+1+1}$ &\!\!\!\!\!\!\!  \!\!\!\! $\so(4)$ & $3$  \\
\!\!$\mathbf{2+2+1+1+1}$ &\!\!\!\!\!\!\!  \!\!\!\! $\spl(2)\oplus\so(3)$ & $4$ \\
\!\!\!$\mathbf{1+1+1+1+1+1+1}$ &\!\!\!\!\!\!\!   \!\!\!\!$\so(7)$ & $5$  
\end{tabular}
\ee
Here the first column gives the $\su(2)$ representations, and the second column gives the unbroken gauge symmetry at the classical level. Finally, the third column gives the number of vacua in the quantum theory. The entries marked `massless' refer to the vacua that classically have abelian factors in the gauge group. These vacua have massless modes also in the quantum theory. As another example, we find for $\Sp(6)$:
\be \label{sp6}
\begin{tabular}{ccc}
 \underline{$\su(2)$ irreps} & \underline{unbroken symmetry} & \underline{number of quantum vacua}  \\
 $\mathbf{6}$ & $\emptyset$ & $1$ \\
 $\mathbf{4+ 2}$ & $\emptyset$ & $1$ \\
 $\mathbf{4+1+1}$ & $\spl(2)$ & $2$  \\
$\mathbf{3+3}$ & $\spl(2)$ & $2$  \\
$\mathbf{2+2+2}$ & $\so(2)$ & $2$ \\
$\mathbf{2+2+1+1}$ & $\so(2)\oplus\spl(2)$ &$\mathrm{massless }$ $(2)$  \\
$\mathbf{2+1+1+1+1}$ & $\spl(4)$ & $3$  \\
$\mathbf{1+1+1+1+1+1}$ & $\spl(6)$ & $4$ 
\end{tabular}
\ee
From the above tables, we see that the number of (massive) quantum vacua is $15$ in both cases, which constitutes a check of S-duality as explained earlier. A few comments are in order. If the unbroken gauge group at the classical level is given by a  product of simple groups then the total number of quantum vacua is given by the product of the dual Coxeter numbers for each factor. However, there is one subtle point in the above analysis: since $\so(4)\cong \su(2)\oplus\su(2)$ one would expect four vacuum states in the $\so(4)$  entry in (\ref{so7}). However, we saw above that only three were required in order for S-duality to work. The explanation is presumably that the gauge group is not $\SO(4) \cong [\SU(2)\otimes\SU(2)]/\ZZ_2$ but really $\mathrm{O}(4)$. The extra discrete gauge symmetry projects out one of the four states leaving three (see e.g.~\cite[p. 32] {Keurentjes:2000b} for a similar discussion in a different context). 

The entries marked massless in the above tables have massless modes (abelian $\ul(1)$'s). One can also contemplate counting such vacua in the following sense (a similar idea  was put forth in \cite{Henningson:2007}). If the classical gauge symmetry is of the form $\ul(1)^l\oplus s$ where $s$ is semi-simple, then the number of quantum vacua for the ``transverse'' part controlled by $s$, leads to a number of `continua' of dimension $l$, i.e.~there is a discrete number of vacua with an $\ul(1)^l$ symmetry. From the above tables, we see that also this generalised counting works for the two theories. 

We could in principle continue the above reasoning to higher ranks. However, it is clear that this method quickly becomes very cumbersome. 

To make progress we will instead use a technique common in the theory of partitions. We are actually not interested in the precise number of vacua in the two theories, we only want to know if their numbers agree. The strategy is to write a generating function for the number of vacua in the two theories and then compare these functions. (A similar technique was employed in \cite{Henningson:2007}.)

From the block-diagonal nature of the $\phi^i$ (\ref{Ti}) it follows that we can construct the 
 generating function in steps. 
Starting with the $\Sp(2n)$ theory we first focus on the odd-dimensional irreps. The contribution to the generating function from $2m$ $(2k{-}1)$-dimensional blocks is $(m{+}1)q^{2m(2k-1)}$, where $q$ is a (formal) variable and $m{+}1$ arises from the dual Coxeter number of $\spl(2m)$ and $2m(2k{-}1)$ is the dimension of the $2m$ blocks. It is clear that we should sum over $m$ ($m$ is fixed in any given vacuum, but can take any value) and take the product over $k$ (all $k$ are allowed, but each can appear at most once). This reasoning leads to:
\be
\prod_{k=1}^{\infty} \sum_{m=0}^{\infty} (m{+}1)\,q^{2m(2k-1)} = \prod_{k=1}^{\infty}\frac{1}{(1-q^{4k-2})^2}\,.
\ee
For the even-dimensional irreps the situation is slightly more involved. 
For $m$ $2k$ dimensional representations with $m> 4$ we get a contribution $(m-2)q^{2mk}$. However, for lower ranks the number of vacua is not equal to $(m-2)$. For $m=4$ we get $3q^{8k}$ (cf.~discussion above), for $m=3$ we get $2q^{6k}$ (since $\so(3)\cong \su(2)$) and for $m=1$ (completely broken gauge symmetry) we get $q^{2k}$. When $m=2$ we get an abelian factor $\so(2)\cong \ul(1)$. To count also the number of vacua with $l$ abelian factors we introduce a new variable $y$ which counts the number of $\ul(1)$'s. For $m=2$ we then get a contribution $yq^{4k}$. Collecting these results, we find
\bea
& & 1+q^{2k} + y q^{4k} + 2q^{6k} + 3q^{8k} + \sum_{m=5}^{\infty} (m-2)q^{2mk} \non \\
&=&\frac{(1-q^{2k}-(1-y)q^{4k} + (3-2y)q^{6k} - (1-y)q^{8k} - q^{10k} + q^{12k})}{(1-q^{2k})^2 }\,.
\eea
Combining the above two expressions we arrive at the result that the coefficient in front of $q^{2n}$ in the power series expansion of 
\be \label{fsp}
\prod_{k=1}^{\infty}\frac{(1-q^{2k}-(1-y)q^{4k} + (3-2y)q^{6k} - (1-y)q^{8k} - q^{10k} + q^{12k})}{(1-q^{2k})^2(1-q^{4k-2})^2 }\,,
\ee
gives the number of (massive) quantum vacua in the $\Sp(2n)$ theory, and the coefficient in front of $y^lq^{2n}$ gives the number of vacua with an $\ul(1)^l$ symmetry. For instance, one can check that the coefficient in front of $q^6$ is 15 and the coefficient on front of $yq^6$ is two, in agreement with the above counting (\ref{sp6}). The number of quantum vacua in the $\SO(n)$ theories can be deduced from (\ref{fsp}) by simply replacing $q^{2k} \leftrightarrow q^{2k-1}$ in this expression (this follows from the results given in the table in section 2). To obtain the generating function for only the $\SO(2n+1)$ theories we need to remove the even part of the function by hand. If S-duality is to hold, the resulting expression should then agree with (\ref{fsp}) multiplied by an additional $q$. Explicitly this requires:
\bea \label{result}
&&\!\!\!q \prod_{k=1}^{\infty} \frac{(1-q^{2k}-(1-y)q^{4k} + (3-2y)q^{6k} - (1-y)q^{8k} - q^{10k} + q^{12k}) }{ (1-q^{2k})^2 (1-q^{4k-2} )^2} \stackrel{?}{=} \\
&&\!\!\!\half \prod_{k=1}^{\infty} \frac{(1-q^{2k-1}-(1{-}y)q^{4k-2} + (3{-}2y)q^{6k-3} - (1{-}y)q^{8k-4} - q^{10k-5} + q^{12k-6}) }{ (1-q^{2k-1} )^2 (1-q^{4k} )^2} - \non \\
&&\!\!\!\half \prod_{k=1}^{\infty} \frac{(1+q^{2k-1}-(1{-}y)q^{4k-2} - (3{-}2y)q^{6k-3} - (1{-}y)q^{8k-4} + q^{10k-5} + q^{12k-6}) }{ ( 1+q^{2k-1})^2 (1-q^{4k} )^2} \,.\non
\eea
To analyze this rather complicated looking expression we first we note 
that the polynomial 
\be \label{P1}
P(x) = 1-x -(1-y)x^2 + (3-2y)x^3 - (1-y)x^{4} - x^{5} + x^{6}\,,
\ee
is a central constituent of the above expression and satisfies $x^6 P(1/x) = P(x)$. Thus, if $x=\al$ is a root of $P(x)$ then so is $x=1/\al$. This means that $P(x)=\prod_{i=1}^3 (x+\al_i )(x+1/\al_i) $ which can be written as 
\be \label{P2}
P(x) = 1 + (\si_1 + \frac{\si_2}{\si_3})x + (\frac{\si_1}{\si_3} +\frac{\si_1\si_2}{\si_3} +\si_2 )x^2 + (\frac{1}{\si_3}+\si_3+\frac{\si_1^2}{\si_3}+\frac{\si_2^2}{\si_3} ) x^3 + \ldots
\ee
where $\si_1 = \al_1+\al_2+\al_3$ and $ \si_2 = \al_1\al_2+\al_2\al_3+\al_3\al_1$ and $\si_3 =\al_1\al_2\al_3$. By identification of (\ref{P1}) with (\ref{P2}) we find in particular that 
\bea \label{sieq}
0 &=& 2(\frac{\si_1}{\si_3} {+}\frac{\si_1\si_2}{\si_3} {+}\si_2 ) + (\frac{1}{\si_3}{+}\si_3+\frac{\si_1^2}{\si_3}{+}\frac{\si_2^2}{\si_3} ) - 1 -2 (\si_1 {+} \frac{\si_2}{\si_3}{+}1) - (\si_1 {+} \frac{\si_2}{\si_3})^2 +1 \non \\
&=& (\si_3-1)\frac{(\si_3^2-2\si_1\si_3+2\si_2\si_3-\si_1^2\si_3 -\si_3+\si_2^2)}{\si_3^2}\,.
\eea
This result means that we can choose the $\al_i$ in such a way that $\si_3=1$, i.e.~$\al_3=\frac{1}{\al_1\al_2}$\footnote[2]{The three solutions to $\si_3^2-2\si_1\si_3+2\si_2\si_3-\si_1^2\si_3 -\si_3+\si_2^2=0$ in (\ref{sieq}) are $\al_1=\al_2 \al_3$ et cycl and therefore lead to the same result, since both $\al_i$ and $1/\al_i$ are roots of $P(x)$.}. After making this choice we have $y=\si_1\si_2$, and $\si_1+\si_2=-1$ i.e.
\be \label{alid}
\al_1 + \al_2+\frac{1}{\al_1\al_2} + \frac{1}{\al_1} + \frac{1}{\al_2} + \al_1\al_2 = -1\,.
\ee
Let us now consider the following identity
\bea \label{ram}
&&(-c,-ac,-bc,-abc,-q^2/c,-q^2/ac,-q^2/bc,-q^2/abc;q^2)_{\infty} \non \\
&&-\, (c,ac,bc,abc,q^2/c,q^2/ac,q^2/bc,q^2/abc;q^2)_{\infty} \\ 
&=&
 2 c(-a,-b,-abc^2,-q^2/a,-q^2/b,-q^2/abc^2,-q^2,-q^2;q^2)_{\infty} \non
\eea
where $(a_1,\ldots,a_i;q^2)_{\infty} = (a_1;q^2)_{\infty} \cdots(a_i;q^2)_{\infty}$ and $(a;q^2)_{\infty} = \prod_{k=0}^{\infty} (1 - a q^{2k})$.
This identity was first written down by Ramanujan in his notebooks in a slightly different form \cite[p. 47, Corollary]{Berndt:1991} (the equivalence between the two expressions was shown in \cite{Baruah:2006}). In the form we have written it here it was recently rediscovered by Warnaar \cite{Warnaar:2005}, who also supplied three different proofs. 

If we set $c=q$, $a=\al_1$ and $b=\al_2$ in (\ref{ram}) and use Euler's famous identity 
$(-q^2;q^2)_{\infty} = (q^2;q^4)_{\infty}^{-1}$ together with $(\pm q;q^2)_{\infty} = (q^2;q^2)_{\infty} (q^4;q^4)^{-1}_{\infty} (\mp q;q^2)^{-1}_{\infty}$ and 
$ (\beta q^2;q^2)_{\infty} = (1-\beta)^{-1}(\beta;q^2)_{\infty} $  we find
\bea
&& \prod_{k=0}^{\infty} \frac{(\al_1 {+} q^{2k+1})(\al_2 {+} q^{2k+1})(\al_1\al_2 {+} q^{2k+1})(\frac{1}{\al_2} {+} q^{2k+1})(\frac{1}{\al_2} {+} q^{2k+1})(\frac{1}{\al_1\al_2} {+} q^{2k+1})}{(1-q^{2k+1})^2(1-q^{4k+2})^2} \non \\
&&\!\!\!\!\! - \prod_{k=0}^{\infty} \frac{(\al_1 {-} q^{2k+1})(\al_2 {-} q^{2k+1})(\al_1\al_2 {-} q^{2k+1})(\frac{1}{\al_2} {-} q^{2k+1})(\frac{1}{\al_2} {-} q^{2k+1})(\frac{1}{\al_1\al_2} {-} q^{2k+1})}{(1+q^{2k+1})^2(1-q^{4k+2})^2} \non \\
&&  = (1+\al_1\al_2)^{-1} (1+\frac{1}{\al_1})^{-1}  (1+\frac{1}{\al_2})^{-1} \times      \\
&& 2q\prod_{k=0}^{\infty} \frac{(\al_1 + q^{2k})(\al_2 + q^{2k})(\al_1\al_2 + q^{2k})(\frac{1}{\al_2} + q^{2k})(\frac{1}{\al_2} + q^{2k})(\frac{1}{\al_1\al_2} + q^{2k})}{(1-q^{2k})^2(1-q^{4k})^2}\,. \non 
\eea
But,
\be 
(1+\al_1\al_2) (1+\frac{1}{\al_1})  (1+\frac{1}{\al_2}) = 2 + \al_1+\al_2 + \al_1\al_2 +\frac{1}{\al_1} +\frac{1}{\al_2} +\frac{1}{\al_1\al_2} = 1 \,,
\ee
using (\ref{alid}). After this observation, together with letting $k \rar k-1$ in the first two products, we find precise agreement 
with (\ref{result}). This concludes the proof that the number of quantum vacua in the $\cN=1^*$ theories with gauge groups $\Sp(2n)$ and $\SO(2n{+}1)$ agree, and provides non-trivial support for S-duality.

\section*{Acknowledgements}
We would like to thank Steve Naculich and Howard Schnitzer for collaboration on \cite{Naculich:2001} which provided the foundation for the present work.  
This work was supported by a grant from the Swedish Science Council.

\begingroup\raggedright\endgroup

\end{document}